\newcommand\relativepath{}

\documentclass[11pt, a4paper]{article}
\usepackage{fullpage}

\newcommand\bigone[1]{}
\newcommand\smallone[1]{#1}
\usepackage{amsfonts,amssymb,amsmath,amsthm,latexsym}
\usepackage{array,xspace}

\newcommand{\ignore}[1]{}

\newcommand{\eps}{\varepsilon}

\newcommand{\etal}{{\em et al.}\xspace}

\def\makeletter#1{%
\expandafter \newcommand \csname b#1\endcsname {\mathbb{#1}}%
\expandafter \newcommand \csname c#1\endcsname {\mathcal{#1}}%
\expandafter \newcommand \csname t#1\endcsname {\widetilde{#1}}%
\expandafter \newcommand \csname ct#1\endcsname {\widetilde{\mathcal{#1}}}%
}
\def\makeletters(#1#2){\makeletter#1\ifx#2.\else\makeletters(#2)\fi}
\makeletters(QWERTYUIOPASDFGHJKLZXCVBNM.)

\def\makeSkob#1#2#3{%
\def\LLL{\left} \def\RRR{\right}
\expandafter \edef \csname #1\endcsname #2##1#3{\SkobInner}
\def\LLL{\bigl} \def\RRR{\bigr}
\expandafter \edef \csname #1A\endcsname #2##1#3{\SkobInner}
\def\LLL{\Bigl} \def\RRR{\Bigr}
\expandafter \edef \csname #1B\endcsname #2##1#3{\SkobInner}
\def\LLL{\biggl} \def\RRR{\biggr}
\expandafter \edef \csname #1C\endcsname #2##1#3{\SkobInner}
\def\LLL{\Biggl} \def\RRR{\Biggr}
\expandafter \edef \csname #1D\endcsname #2##1#3{\SkobInner}
\def\LLL{} \def\RRR{}
\expandafter \edef \csname #1O\endcsname #2##1#3{\SkobInner}
}

\def\SkobInner{\LLL(##1\RRR)} \makeSkob{s}[]
\def\SkobInner{\LLL[##1\RRR]} \makeSkob{sk}[]
\def\SkobInner{\LLL\lbrace##1\RRR\rbrace} \makeSkob{sfig}{}{}
\def\SkobInner{\LLL\lfloor##1\RRR\rfloor} \makeSkob{floor}[]
\def\SkobInner{\LLL\lceil##1\RRR\rceil} \makeSkob{ceil}[]
\def\SkobInner{\LLL\langle##1\RRR\rangle} \makeSkob{ip}<>
\def\SkobInner{\LLL|##1\RRR\rangle} \makeSkob{ket}|>
\def\SkobInner{\LLL|##1\RRR|} \makeSkob{abs}||
\def\SkobInner{\LLL\|##1\RRR\|} \makeSkob{norm}||
\def\SkobInner{\LLL\|##1\RRR\|_{\noexpand\mathrm F}} \makeSkob{normFrob}||
\def\SkobInner{\LLL\|##1\RRR\|_{\noexpand\mathrm{tr}}} \makeSkob{normtr}||

\def \elem[#1]{[\![#1]\!]}
\def \bigfrac#1/{\left.#1\right/}
\def \bigfracR/#1.{\left/#1\right.}

\newcommand{\pfstart}{\begin{proof}} 
\newcommand{\pfsketch}{\begin{proof}[Proof sketch]}
\newcommand{\pfend}{\end{proof}} 
\newcommand{\itemstart}{\begin{itemize}\itemsep0pt}
\newcommand{\itemend}{\end{itemize}}
\newcommand{\descrstart}{\begin{description}\itemsep0pt}
\newcommand{\descrend}{\end{description}}
\newcommand{\enumstart}{\begin{enumerate}\itemsep0pt}
\newcommand{\enumend}{\end{enumerate}}

\newcommand{\CrossRef}[2]{#1~\ref{#2}}

\bigone{\newtheorem{thm}{Theorem}[chapter]}
\smallone{\newtheorem{thm}{Theorem}}

\newcommand{\maketheorem}[2]{
\newtheorem{#1}[thm]{#2}
\expandafter\def \csname ref#1\endcsname ##1{\CrossRef{#2}{#1:##1}}
}

\maketheorem{lem}{Lemma}
\maketheorem{prp}{Proposition}
\maketheorem{cor}{Corollary}
\maketheorem{clm}{Claim}
\maketheorem{fact}{Fact}

\theoremstyle{definition}
\maketheorem{rem}{Remark}
\maketheorem{obs}{Observation}
\maketheorem{defn}{Definition}
\maketheorem{exm}{Example}
\maketheorem{assump}{Assumption}

\def \rf(#1:#2){\csname ref#1\endcsname{#2}}
\def \rfitem(#1@#2){\rf(#2)(\ref{#1@#2})}

\def\mycommand#1#2{
\expandafter\newcommand \csname#1\endcsname {#2}%
}
\def\remycommand#1#2{
\expandafter\renewcommand \csname#1\endcsname {#2}%
}

\newcommand\draft[1]{}
\newcommand\release[1]{#1}
\release{\usepackage[breaklinks]{hyperref}}

\title{Quantum Algorithm for Monotonicity Testing on the Hypercube}
\author{
Aleksandrs Belovs
\thanks{Faculty of Computing, University of Latvia.}
 \and 
Eric Blais
\thanks{David R. Cheriton School of Computer Science, University of Waterloo.}
}
\date{}

\begin{document}
\maketitle

\mycommand{bool}{\{0,1\}}
\mycommand{cube}{\bool^n}

\begin{abstract}
In this note, we develop a bounded-error quantum algorithm that makes $\tO(n^{1/4}\eps^{-1/2})$ queries to a function $f\colon \cube\to\bool$, accepts a monotone function, and rejects a function that is $\eps$-far from being monotone.
This gives a super-quadratic improvement compared to the best known randomized algorithm for all $\eps = o(1)$.  The improvement is cubic when $\eps = 1/\sqrt{n}$.

\end{abstract}

\section{Introduction}
The problem of testing monotonicity of Boolean functions is one of the fundamental---and most extensively studied---problems in property testing.
Let $\preceq$ denote the bitwise partial order on the Boolean hypercube $\cube$, i.e., $x\preceq y$ iff $x_j\le y_j$ for all $j\in[n]$.
The function $f\colon\cube\to\bool$ is {\em monotone} iff $f(x)\le f(y)$ for all $x\preceq y$, and it is {\em $\eps$-far from monotone} if it cannot be made monotone by changing its value on at most $\eps2^n$ inputs.  An {\em $\eps$-tester for monotonicity} is a randomized algorithm that distinguishes monotone functions from those that are $\eps$-far from monotone with large (say, $\frac23$) probability.

The study of the monotonicity testing problem was initiated by Goldreich \etal~\cite{goldreich:monotonicityTestingNew}, who showed that it is possible to $\eps$-test monotonicity of functions $f\colon \cube\to\bool$ with $O(n/\eps)$ queries to $f$.
Their algorithm is called the {\em edge tester}, and it is very simple:  Repeatedly sample edges of the hypercube uniformly at random, and test whether the values of $f$ on their endpoints violate monotonicity.
The example of an anti-dictator function, $f(x) = \neg x_i$, shows that the analysis of this algorithm is tight.

The edge tester remained the most efficient monotonicity testing algorithm for more than a decade, until Chakrabarti and Seshadhri~\cite{chakrabarty:sublinearMonotonicity} introduced a new $\eps$-tester for monotonicity that requires only $\tO(n^{7/8}\eps^{-3/2})$ queries.
The Chakrabarti--Seshadhri algorithm, like the edge tester, is a \emph{pair tester}:  It repeatedly picks pairs of inputs $x \preceq y$ from the hypercube and verifies that $f(x) \le f(y)$ on each pair. Unlike the edge tester, however, the Chakrabarti--Seshadhri algorithm selects pairs of inputs that have Hamming distance up to $O(\sqrt{n})$.
The same high-level approach has since been used by Chen, Servedio, and Tan~\cite{chen:newAlgorithmsLowerBoundsMonotonicity} to obtain a $\tO(n^{5/6}\eps^{-4})$-query $\eps$-tester for monotonicity and, very recently, by Khot, Minzer, and Safra in their beautiful paper~\cite{khot:monotonicityTesting} which shows that $\tO(\sqrt{n}/\eps^2)$ queries suffice to $\eps$-test monotonicity.

In summary, this line of research had led to two incomparable $\eps$-testers for monotonicity: the edge tester with query complexity $O(n/\eps)$, and the Khot--Minzer--Safra algorithm with query complexity $\tO(\sqrt{n}/\eps^2)$. 
The first one has better dependence on $\eps$, the second one on $n$.
Furthermore, these two algorithms---along with every other algorithm that has been proposed for testing  monotonicity of Boolean functions---are pair testers. Let us also note that pair testers are \emph{non-adaptive} algorithms (they can select all their queries in advance) and have \emph{one-sided error} (always accept a monotone function).

In this paper, we consider the query complexity of \emph{quantum} algorithms that test monotonicity of Boolean functions. See~\cite{montanaro:quantumProperyTest} for a recent survey on quantum property testing.
While the quantum query complexity of the monotonicity testing problem has not been explicitly studied before, we can apply
 quantum amplitude amplification~\cite{brassard:amplification} to obtain a quadratic improvement on the query complexity of any pair tester. As a result, the edge tester and Khot--Minzer--Safra algorithms imply that we can $\eps$-test monotonicity with $O(\sqrt{n/\eps})$ and $\tO(n^{1/4}/\eps)$ quantum queries, respectively.
Our main result is a simple quantum algorithm that combines the best dependence from both of these algorithms.

\begin{thm}
\label{thm:main}
It is possible to $\eps$-test monotonicity of a function
$f\colon\cube\to\bool$ with $O(\frac1{\sqrt{\eps}} \cdot n^{1/4} \log n)$ quantum queries.
\end{thm}

Let us compare this result with known lower bounds on classical algorithms.  Lower bounds for this problem are notoriously hard.  For many years, the best known lower bound on the {\em non-adaptive} randomized query complexity was $\Omega(\log n)$ for constant $\eps$ by Fischer \etal~\cite{fischer:monotonicitytesting}.  Recently, it was improved by Chen \etal~\cite{chen:newAlgorithmsLowerBoundsMonotonicity, chen:booleanMonotonicityRequiresSqrtn} to almost $\Omega(\sqrt{n})$, essentially matching the query complexity of the Khot--Minzer--Safra algorithm.  The best known {\em adaptive} lower bound is only $\Omega(\log n)$, which immediately follows from the non-adaptive lower bound.

For pair testers, more lower bounds are known.  First, Bri{\"e}t \etal~\cite{briet:monotonicityAndRouting} proved that any pair tester with query complexity of the form $O(\alpha(n)/\eps)$ must have $\alpha(n) = \Omega(n/\log n)$.
Also, Khot, Minzer, and Safra~\cite{khot:monotonicityTesting} gave an example of a family of functions that are at distance $\Theta(1/\sqrt{n})$ to being monotone, but for which any pair tester needs $\Omega(n^{3/2})$ queries to find a non-monotone pair with constant probability.  Our algorithm, on the other hand, needs only $\tO(\sqrt{n})$ queries, which constitutes a cubic improvement.

This is an interesting development, as very few super-quadratic but still polynomial speed-ups are known.  We can only mention a cubic speed-up for exponential congruences by van Dam and Shparlinski~\cite{vanDam:exponentialCongruences}, and quartic speed-ups for finding counterfeit coins by Iwama~\etal~\cite{iwama:quantumCounterfeit}, and learning the ``exactly-half junta'' by Belovs~\cite{belovs:learningSymmetricJuntas}.

Our algorithm is based on technical results from~\cite{khot:monotonicityTesting} and the (dual) adversary bound.
The latter characterises quantum query complexity up to a constant factor, as shown by Reichardt \etal~\cite{reichardt:spanPrograms,lee:stateConversion}.
Previously, the adversary bound was used for formula evaluation~\cite{reichardt:formulae, zhan:treesWithHiddenStructure}, triangle and other subgraph detection~\cite{belovs:learning, lee:learningTriangle, belovs:learningClaws}, the $k$-distinctness problem~\cite{belovs:learningKDist}, and learning symmetric juntas~\cite{belovs:learningSymmetricJuntas}.  
This paper demonstrates an application to property testing.

\section{Adversary Bound}
In this section, we introduce the dual adversary bound.
That is the main tool we use in the construction of our algorithm.  We use it to distinguish total Boolean functions, so let us define the version of the bound tailored for this special case.  

Let $n$ be a positive integer, and assume that $\cX$ and $\cY$ are two disjoint sets of total functions from $\{0,1\}^n$ to $\{0,1\}$.  Let us also denote $\cD = \cX\cup\cY$.
We will deal with the following problem.  Given a query access to a function $f\in\cX\cup\cY$, the task is to detect whether $f\in\cX$ or $f\in\cY$.  
For this problem, the (dual) adversary bound is equal to the optimal value of the following optimization problem:
\begin{subequations}
\label{eqn:advDual}
\begin{alignat}{3}
&\mbox{\rm minimize} &\quad& \max_{f\in \cD}\sum\nolimits_{z\in \{0,1\}^n} X_z\elem[f,f]  \label{eqn:advDualObjective} \\
& \mbox{\rm subject to}&& \sum\nolimits_{z: f(z)\ne g(z)} X_z\elem[f, g] = 1 &\quad& \text{\rm for all $f\in\cX$ and $g\in\cY$;} \label{eqn:advDualCondition} \\
&&& X_z\succeq 0 && \mbox{\rm for all $z\in \{0,1\}^n$,} \label{eqn:advDualSemidefinite}
\end{alignat}
\end{subequations}
where $X_z$ are $\cD\times\cD$ positive semi-definite matrices.  The adversary bound is very useful because of the following result:

\begin{thm}[\cite{hoyer:advNegative, lee:stateConversion}]
\label{thm:adv}
The quantum query complexity of distinguishing $\cX$ and $\cY$ is equal to the value of the adversary bound~\rf(eqn:advDual), up to a constant factor.
\end{thm}

In particular, Theorem~\ref{thm:adv} implies that the value of every feasible solution to the optimization problem~\rf(eqn:advDual) gives an upper bound on the quantum query complexity of the corresponding problem.  

\section{Proof of Theorem~\ref{thm:main}}
\mycommand{fin}{f^{-1}(0)}
\mycommand{fip}{f^{-1}(1)}

We complete the proof of Theorem~\ref{thm:main} by constructing a feasible solution to the adversary bound~\rf(eqn:advDual) in the case where $\cX$ is the set of all monotone functions and $\cY$ is the set of all functions that are $\eps$-far away from any monotone function.

Let us introduce some additional notation.
An {\em edge} of the hypercube is a pair $xy$, where $x\prec y$ and $x$ and $y$ differ in exactly one position.
For a fixed function $f\colon\cube\to\bool$, an $ab$-edge is an edge $xy$ such that $f(x)=a$ and $f(y)=b$.
The {\em total influence} or {\em average sensitivity} $\bI(f)$ is the number of edges $xy$ of the hypercube such that $f(x)\ne f(y)$, divided by $2^{n-1}$.  For a monotone function, it is known to be $O(\sqrt{n})$~\cite[Theorem 2.33]{odonnell:analysis}.

Unlike the Khot--Minzer--Safra algorithm, which tests pairs of inputs at significant distance, our algorithm is essentially the edge tester.  
Our main contribution is the optimization of the $O(\sqrt{n/\eps})$ complexity, which we would get with the straightforward implementation, as mentioned in the introduction.
For that, we need the following technical result:

\begin{lem}[{\cite[Lemma 7.1]{khot:monotonicityTesting}}]
\label{lem:tech}
For every $f\colon\cube\to\bool$ that is $\eps$-far from being monotone,
there exists a bipartite graph $G_f = (V,W,E)$ satisfying the following properties:
\itemstart
\item the parts of the graph satisfy $V\subseteq \fip$ and $W\subseteq\fin$;
\item every edge $e\in E$ is also a $10$-edge of the hypercube;
\item the graph has $|E| = \Omega(\eps 2^n \sqrt{\Delta(G_f)} /\log^2 n)$ edges, where $\Delta(G_f)$ is the maximal degree of $G(f)$.
\itemend
\end{lem}

For each $f\in\cY$, let us fix a graph $G_f$ as in \rf(lem:tech).  Let $E_f$ be the set of edges of $G_f$, and $\deg_f(x)$ be the degree of a vertex $x$ in $G_f$.

We construct a feasible solution to~\rf(eqn:advDual) in the form $X_x = Y_x + Z_x$.  In the following, $K$ and $L$ are some constants that we will define later.  The matrix $Y_x$ is defined by $\psi_x\psi_x^*$, where
\[
\psi_x\elem[f] = 
\begin{cases}
1/\sqrt{K},& f\in\cX; \\
\sqrt{K}\deg_f(x)/|E_f|,& f\in\cY.
\end{cases}
\]
The matrix $Z_{x}$ is defined as $\sum_{j\in [n]}Z_{x,j}$, where $Z_{x,j} = \phi_{x,j}\phi_{x,j}^*$ with
\[
\phi_{x,j}\elem[f] = 
\begin{cases}
-1/\sqrt{L},& \text{$f\in\cX$, $x_j=0$, and $(x,x^{\oplus j})$ is a $01$-edge;} \\
\sqrt{L}/|E_f|,& \text{$f\in\cY$, $x_j=0$, and $(x,x^{\oplus j})\in E_f$;} \\
0,& \text{otherwise;}
\end{cases}
\]
where $x^{\oplus j}$ stands for the string $x$ with the $j$th bit flipped.

Let us check that this satisfies~\rf(eqn:advDualCondition).
Fix two functions $f\in\cX$ and $g\in\cY$.  
It is not hard to see that
\[
\sum_{x\colon f(x)\ne g(x)} Y_x\elem[f,g] = \sum_{xy\in E_g} \frac1{|E_g|} \sB[{\skA[f(x)\ne g(x)]+\skA[f(y)\ne g(y)]}],
\]
where $[P]=1$ iff the proposition $P$ is true, and $[P]=0$ otherwise.  On the other hand,
\[
\sum_{x\colon f(x)\ne g(x)} Z_x\elem[f,g] = -\sum_{xy\in E_g} \frac1{|E_g|} \skB[ (f(x)\ne g(x))\wedge (f(y)\ne g(y)) ] .
\]
As for any $xy\in E_g$, at least one of the conditions $f(x)\ne g(x)$ or $f(y)\ne g(y)$ is satisfied, the last two equations prove~\rf(eqn:advDualCondition).

Let us estimate the objective value~\rf(eqn:advDualObjective) of this solution.	 For a function $f\in\cX$, we have
\begin{equation}
\label{eqn:XObj}
\sum_{x\in\cube} X_x\elem[f,f] = 2^n \sB[\frac1K + \frac{\bI(f)}{2L}] = 2^n\cdot O\sB[\frac1K + \frac{\sqrt{n}}L].
\end{equation}
Meanwhile, for $g\in\cY$, we have
\begin{align}
\sum_{x\in\cube} X_x\elem[g,g] &=  \sum_{x\in\cube} \frac{K \deg_g(x)^2}{|E_g|^2} + |E_g|\cdot\frac{L}{|E_g|^2}.\notag 
\end{align}
Using the inequality $\sum_x \deg_g(x)^2 \le \Delta(G_g) \sum_x \deg_g(x)$ and the identity $\sum_x \deg_g(x) = 2|E_g|$, we observe that
\begin{align}
\sum_{x\in\cube} X_x\elem[g,g] 
&\le K \frac{\Delta(G_g)}{|E_g|}\cdot \frac{\sum_x \deg_g(x)}{|E_g|} + \frac{L}{|E_g|} \notag 
= 2K \frac{\Delta(G_g)}{|E_g|} + \frac{L}{|E_g|} \notag.
\end{align}
We can now apply Lemma~\ref{lem:tech} to obtain
\begin{align}
\sum_{x\in\cube} X_x\elem[g,g] 
&\le \frac{\log^2 n}{\eps 2^n} \cdot O\s[ K\sqrt{\Delta(G_g)} + L ] 
= \frac{\log^2 n}{\eps 2^n} \cdot O\s[ K\sqrt{n} + L ] . \label{eqn:YObj}
\end{align}

Comparing~\rf(eqn:XObj) and~\rf(eqn:YObj), we see that the objective value is minimized when we set 
 $K = 2^n\sqrt{\eps}n^{-1/4}/\log n$ and $L = 2^n\sqrt{\eps}n^{1/4}/\log n$. Taking these values, 
 the objective value~\rf(eqn:advDualObjective) is $O(n^{1/4}\eps^{-1/2}\log n)$, as desired.

%
%

\section*{Acknowledgements}
A.B. is supported by FP7 FET Proactive project QALGO.
Part of this work was completed while A.B. was at MIT, supported by Scott Aaronson's Alan T. Waterman Award from the National Science Foundation.  A.B. thanks Andris Ambainis for his comments on the previous versions of this note. 

Part of this work was completed E.B.~was a Simons Postdoctoral Fellow at MIT.
\bibliographystyle{\relativepath habbrvM}
\bibliography{../../bib} 

\end{document}